\documentclass[10pt,conference]{IEEEtran}
\pdfoutput=1

\usepackage{xcolor}
\usepackage{booktabs}
\usepackage{fontawesome}
\usepackage{subfigure}
\usepackage{siunitx}
\usepackage{bm}
\usepackage{xspace}
\usepackage{float}
\usepackage{graphicx}
\usepackage{accents}
\usepackage{threeparttable}
\usepackage{diagbox}
\usepackage{textcomp}
\usepackage[skins]{tcolorbox}
\usepackage{enumitem}
\usepackage{graphicx}
\usepackage{multirow}
\usepackage{verbatim}
\usepackage{longtable}
\usepackage{pgf-pie} 
\usepackage{cite}
\usepackage{amssymb,amsfonts}
\usepackage{algorithmic}
\usepackage{algorithm}
\usepackage{xurl}
\usepackage{ragged2e}
\usepackage[hidelinks]{hyperref}

\def\BibTeX{{\rm B\kern-.05em{\sc i\kern-.025em b}\kern-.08em
    T\kern-.1667em\lower.7ex\hbox{E}\kern-.125emX}}

\newlength{\fsize}

\tcbset{
  my box/.style={
    enhanced,
    colframe=#1!80,
    colback=#1!10,
    attach boxed title to top left={xshift=0.2cm, yshift=-0.2cm},
    boxed title style={
      colback=#1!80,
      outer arc=0pt,
      arc=0pt,
      top=0pt,
      bottom=0pt,
    },
  },
}
\newtcolorbox{result-rq}[1]{
  my box=black,
  title=#1,
  boxrule=1.2pt,top=6pt,bottom=3.5pt,left=6pt,right=6pt
}

\begin{document}
\title{Revealing the value of Repository Centrality in lifespan prediction of Open Source Software Projects}


\author{
   \IEEEauthorblockN{Runzhi He\IEEEauthorrefmark{1}\IEEEauthorrefmark{2}, Hengzhi Ye\IEEEauthorrefmark{1}\IEEEauthorrefmark{2}, Minghui Zhou\IEEEauthorrefmark{1}\IEEEauthorrefmark{3}}
    \IEEEauthorblockA{\IEEEauthorrefmark{1}School of Computer Science, Peking University, China\\
    Key Laboratory of High Confidence Software Technologies, Ministry of Education, China\\
    rzhe@pku.edu.cn, yhz20@mails.tsinghua.edu.cn, zhmh@pku.edu.cn}    
}

\maketitle

\begingroup\renewcommand\thefootnote{\IEEEauthorrefmark{2}}
\footnotetext{These authors contributed equally to this work.}
\endgroup

\begingroup\renewcommand\thefootnote{\IEEEauthorrefmark{3}}
\footnotetext{Minghui Zhou is the corresponding author.}
\endgroup

\thispagestyle{plain}\pagestyle{plain}
\begin{abstract}
\textit{\normalfont\textit{ Background:}}
Open Source Software (OSS) is the building block of modern software. However, the prevalence of project deprecation in the open source world weakens the integrity of the downstream systems and the broad ecosystem. 
Therefore it calls for efforts in monitoring and predicting project deprecations, empowering stakeholders to take proactive measures.
 {\normalfont\textit{Challenge:}}
    Existing techniques mainly focus on static features on a point in time to make predictions, resulting in limited effects.
{\normalfont\textit{Goal:}}
    We propose a novel metric from the user-repository network, and leverage the metric to fit project deprecation predictors and prove its real-life implications.
{\normalfont\textit{Method:}}
   We establish a comprehensive dataset containing 103,354 non-fork GitHub OSS projects spanning from 2011 to 2023. 
   We propose repository centrality, a family of HITS (Hyperlink-Induced Topic Search) weights that captures shifts in the popularity of a repository in the repository-user star network.
   Further with the metric, we utilize the advancements in gradient boosting and deep learning to fit survival analysis models 
   to predict project lifespan or its survival hazard.
{\normalfont\textit{Results:}}
    Our study reveals a correlation between the HITS centrality metrics and the repository deprecation risk. 
        A drop in the HITS weights of a repository indicates a decline in its centrality and prevalence, 
    leading to an increase in its deprecation risk and a decrease in its expected lifespan.
    Our predictive models powered by repository centrality and other repository features achieve satisfactory accuracy on the test set, with repository centrality being the most significant feature among all.
{\normalfont\textit{Implications:}}
    This research offers a novel perspective on understanding the effect of prevalence on the deprecation of OSS repositories. Our approach to predict repository deprecation help detect health status of project and take actions in advance, fostering a more resilient OSS ecosystem.
\end{abstract}

\section{Introduction}
\label{sec:intro}

The prosperous world of modern software does not emerge from nothing. Modern software relies on hundreds, and often thousands, of other pieces of software to function, develop, and maintain. Software engineers refer to the software that others depend on as dependencies. 
Thanks to the power of open source software (OSS) dependencies, 
developers can easily reuse code from various code hosting platforms, such as GitHub. Synopsys conducted an analysis of 1700 codebases from 17 different application domains in their annual Open Source Security and Risk Analysis Report (OSSRAR)~\cite{OSSRAR}. They found that 76\% of the analyzed codebases are open source, and 96\% of them use third-party open-source components.

Open source software, although extremely useful, is often unreliable in the sense that there is no way to know if and when support for the project could cease, i.e., become deprecated. 
Over the years, many influential open source projects have declined, including GitHub's well-known Atom text editor~\cite{atom}, Adobe's HTML editor Brackets~\cite{brackets}, and the JavaScript library faker.js~\cite{faker}, whose deprecation caused chaos. Besides, OSSRAR~\cite{OSSRAR} reveals that over 91\% of the analyzed dependencies have not shown any sign of maintenance in two years. 
Numerous factors contribute to the deaths of open source projects, including maintainers' loss of interest or lack of time, the emergence of more attractive alternatives, and the uncovering of serious security vulnerabilities~\cite{robbes2012developers,sawant2016reaction}.

The impact of repository deprecation extend beyond individual software projects, potentially triggering a domino effect that weakens the foundation of interconnected software supply chain at scale. 
Software systems relying on deprecated repositories are susceptible to security vulnerabilities and compatibility issues, which could instigate cascading failures. 
Moreover, the abrupt deprecation of a repository (e.g. the \textit{left-pad} incident in 2015~\cite{npm_leftpad}) may introduce substantial disruptions in the development lifecycle, forcing developers into ``emergency mode'' and dedicate significant time and effort into the identification and integration of alternatives. 

In light of the potential hazards, the capacity to anticipate repository deprecation emerges as a vital necessity. By accurately predicting the obsolescence of a code repository, developers can undertake proactive measures to counteract the ensuing risks. This might encompass the identification of substitute repositories or packages, strategizing for code migration, or even contributing to the preservation of the repository to avert its deprecation.

To tackle the problem, extensive effort~\cite{khondhu2013all,samoladas2010survival,li2022ossara, valiev2018ecosystem} has been devoted to investigating the factors contributing to repository deprecation. 
However, there is a lack of observation of variations in repository-related features over continuous periods. 
Furthermore, there is a 
room for exploring the
characteristics beyond those directly obtainable to predict repository deprecation,
such as centrality features within the network constructed by the repositories. 
This gap in the existing body of knowledge underscores the need for a more comprehensive and nuanced understanding of these overlooked aspects.  

Therefore, our work endeavors to analyze the popularity dynamics of an OSS repository over time, seeking to predict the prospective lifespan and risk of deprecation of a repository from the current time point. This temporal perspective allows us to capture trends and patterns that could provide more precise predictions of deprecation, enhancing the resilience of the open source ecosystem.

We propose the following research questions (RQs): 

\textbf{RQ1:} Can we model the popularity dynamics of an OSS repository in social coding platform? 

\textbf{Results:} 
 We adopt the HITS algorithm to model
 the ``star'' network of a GitHub repository with  
 a bipartite graph (where users are Hubs and repositories are authorities), and propose repository centrality metric to reflect the popularity of a repository. 
 The metric is a combination of the HITS weight, rank-normalized HITS weight, and z-score normalized HITS weight, providing a nuanced understanding of a repository's position within the open-source ecosystem.

\textbf{RQ2:} How effective is the proposed repository centrality metric in predicting repository deprecation? 

 \textbf{Results:} 
Based on the proposed repository centrality metric, we leverage gradient boosting and deep learning techniques to fit survival analysis models, and to forecast the lifespan and survival hazard of a GitHub repository. The prediction accuracy is high and the proposed metric proves to be effective.


\vspace{.3em}
The contributions of this paper are:
\begin{itemize}
        \item A novel comprehensive metric ``repository centrality'' to capture the shift of project popularity.
    \item A high-performance project decline prediction model that utilizes only open data and has been validated with real-world examples.
    \item A powerful language model to identify real repository deprecation from READMEs and descriptions.
    \item A large-scale and comprehensive dataset with 51,677 validated GitHub project deprecations.
\end{itemize}

We provide a replication package at \url{https://figshare.com/s/981beaa8fc4a93c9c7e0}.

\section{Background and Related Work}\label{backgroundandmoti}

\subsection{Deprecation of OSS}

Deprecation is a prominent problem due to its disruptive nature; it can cause disarray in software development processes, and when developers fail to realize that one of their dependencies is deprecated, it has the potential to introduce security bugs via the software supply chain~\cite{OSSRAR}. Consequently, researchers have tried to understand the implications. To this extent, studies by Robbes et al.~\cite{robbes2012developers} and Sawant et al.~\cite{sawant2016reaction} have laid the groundwork for understanding the nature, causes, and implications of deprecation in OSS. These studies suggest that deprecation can stem from factors, including the loss of interest by maintainers, the emergence of more efficient alternatives, or the unearthing of consequential security vulnerabilities. Subsequent research, such as the study by Kula et al.~\cite{kula2017impact}, indicates that a significant portion of libraries in package ecosystems like npm and RubyGems become deprecated with time and additionally identifies the prevalence of deprecation amongst OSS projects.

To remedy deprecation disruptions, neoteric research has attempted to identify unmaintained projects using advanced machine learning techniques ~\cite{coelho2018identifying, coelho2020github}, as well as employing centrality measures to forecast the trend of package deprecation~\cite{mujahid2021toward}. However, these studies focus on static indicators. 
Our research improves upon the foundational studies by incorporating dynamic factors to offer a more nuanced insight and improved predictive accuracy. We believe this research is valuable as it will help identify deprecation before it happens, allowing developers to plan and implement countermeasures, thereby minimizing and perhaps entirely mitigating risks associated with deprecation in OSS.

\subsection{Network Analysis}\label{relatedwork}

Although a marvel, communities can grow exponentially, quickly becoming an intricate web of networks. Therefore, to facilitate the comprehension of such a complex network, studies have utilized network models to unravel the patterns of such communities~\cite{clauset2004finding}. Similarly, in the software ecosystem, the effectiveness of network analysis has nurtured an understanding of software development attributes such as dependency management and bug proliferation. For instance, one study leveraged network centrality measures to predict the likelihood of future software changes~\cite{pinzger2008can}. Additionally, another study employed network metrics to forecast post-release defects~\cite{nagappan2010change}. However, we observe a noticeable gap in the current literature concerning project deprecation. Firstly, there is a lack of consideration for dynamic factors in the ecosystem. Secondly, there is a lack of utilization of network centrality features within the repository network. Accordingly, we believe that utilizing network analysis will provide a novel perspective in understanding the deprecation of OSS repositories.

\begin{figure*}[htbp]
    \centering
    \includegraphics[width=0.8\linewidth, trim=10 0 0 0, clip]{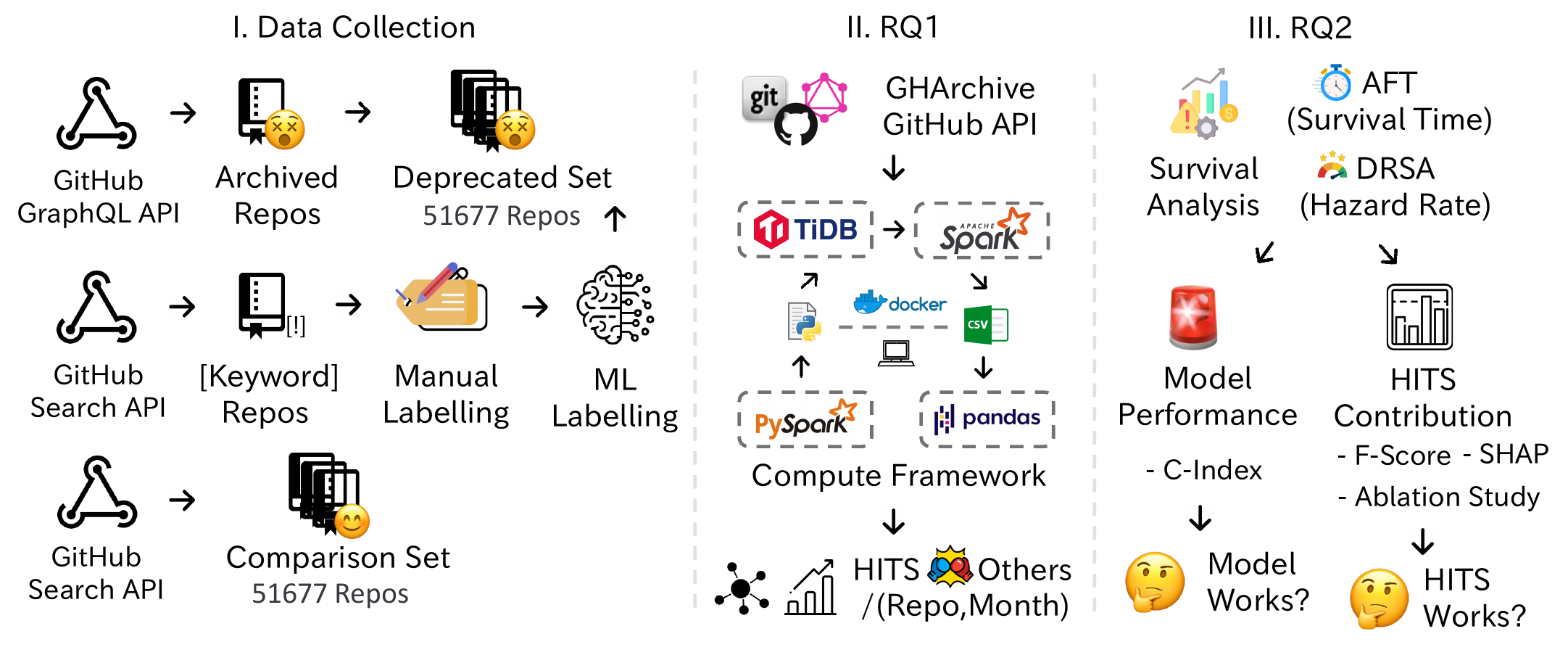}
    \vspace{-1em}
    \caption{Research Framework}
    \label{fig:method}
    \vspace{-1em}
\end{figure*}


\section{Data Collection}\label{empirical}

Figure~\ref{fig:method} presents the methodology of this study. 
In this section we explain how we spend significant effort to construct a comprehensive dataset that is critical to this study and the like, as shown on the left side of Figure~\ref{fig:method}.
For the convenience of operation, we define project deprecation in Section~\ref{ss:difinitiondeprecation}. 
We introduce how we select projects and construct dataset in Section~\ref{ss:projectselection} and ~\ref{ss:dataprocessing} respectively.




\subsection{Definition of Repository Deprecation}\label{ss:difinitiondeprecation}

There is no \textit{de facto} definition of repository deprecation. 
Rather than defining the concept on all repositories in the open source world, we first limit the scope of our research to GitHub. 
GitHub is a code hosting platform of global prominence and extensive use, with a vast array of projects and contributors, ensuring our dataset's scale and diversity.
Besides, GitHub's rich development history and ``archive repository'' feature highlight repository deprecations, ensuring the number and correctness of positive samples.

\begin{figure}
\centering
\subfigure[Archiving a Repository]
{
\centering
\includegraphics[scale=0.35]{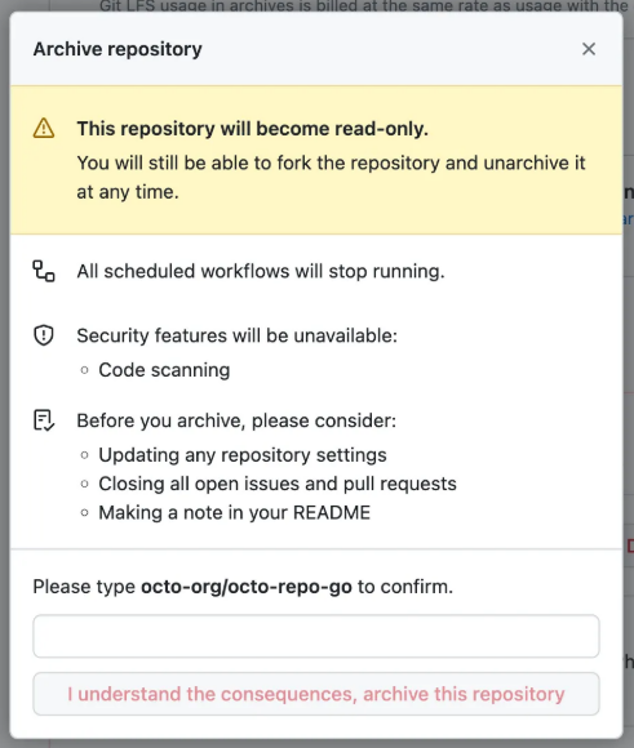}
\label{archive}
}
\subfigure[Deprecation-indicating Keywords]
{
\centering
\includegraphics[scale=0.42]{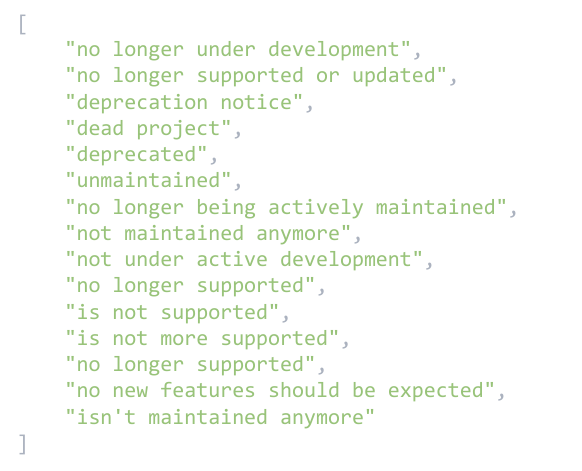}
\label{declaration}
}
\vspace{-.5em}
\caption{Two Methods to Identify Deprecations}
\label{fig:how_to_ident_deprecate}
\vspace{-1em}
\end{figure}

We follow the state of practice of GitHub project maintainers and define a repository to be deprecated if it meets any of the following criteria (shown in Figure~\ref{fig:how_to_ident_deprecate}):


\begin{itemize}
    \item \textbf{Repository is archived.}
GitHub provides a feature for archiving repositories~\cite{GithubArchive}, which is a clear indicator of the discontinuation of maintenance. When a repository is archived, it transitions into a read-only state, prohibiting any further updates or deletions. The process of archiving is straightforward for maintainers, involving simple navigation through ``Settings → Danger Zone → Archive Repository'' on GitHub. This feature has gained considerable traction, and nowadays a considerable number of projects are opting for this approach when deprecating their repositories.
    \item \textbf{Repository has deprecation-indicating keywords in its README or description.}
In noticeable cases, projects opt to announce their deprecation in the repository description or the README file, rather than resorting to the archive feature. (Part of the reason is that the ``archive repository'' feature was not a thing before 2017~\cite{archive_2017}). An example is
the EntityFramework~\cite{ZZProjects}, which communicates its discontinuation by stating ``this library is no longer supported since 2015'' at the outset of its README file. This method of signaling deprecation has also been noted by Coelho et al.~\cite{coelho2017modern} and utilized as a criterion for identifying discontinued repositories.
\end{itemize}

These two methods, shown in Figure~\ref{archive} and~\ref{declaration}, collectively form the basis of our operational definition of deprecation, enabling us to systematically identify and categorize deprecated repositories.

\subsection{Project Selection}\label{ss:projectselection}

Given the vast number of project repositories on GitHub, it is imperative to curate a reasonably sized, clean experimental dataset. This necessitates rigorous selection and aggregation processes for our data sources.

Our initial task involves the construction of a comprehensive dataset of deprecated GitHub repositories, as no existing dataset aptly fits the context of our study, prompting the need for a bespoke dataset. Historically, research in this area has been limited to a narrow scope, focusing primarily on a few top-tier GitHub projects. For instance, the work of Coelho et al.~\cite{coelho2017modern} was confined to failed projects within the top 5000 GitHub projects as ranked by star count. 
The limited sample size is both insufficient for generating robust results from a regression model and vulnerable to validity threats for not considering less popular repositories.
However, researchers limited the scale of the dataset for a reason: to mitigate the threat of false positives, former work involves a considerable number of manual labeling, which translates into hundreds of hours of labor. To address the limitation, we: 1) utilize the ground truth of archived projects on GitHub, which are guaranteed not to be false positives; 2) label
and cross-validate 1,200 samples that declare their deprecation in natural language, and leverage recent advances in language models to train a sentence-transformer-based classifier to label the remaining 20,358.

\emph{Archived Repositories.}
We first use the GitHub GraphQL API~\cite{graphql} to gather all archived GitHub repositories that boast a star count exceeding 32. This threshold is dictated by the constraints of the GitHub GraphQL search API and serves as a filtering criterion to eliminate trivial projects. This star-count-based approach is a common filtering method in software engineering studies, once applied by Xiao et al.~\cite{xiao2022recommending}.

\emph{Repositories with deprecation-indicating keywords.}
Resulting from the combination of historical factors and maintainers' preferences, not all GitHub projects signal the end of life by archiving the repository.
We employ the GitHub GraphQL API to search all non-fork GitHub repositories with the indicating keywords cataloged by Coelho et al.~\cite{coelho2017modern} in their descriptions and README files.
As researchers have found, string matching introduces a unignorable number of false positives (e.g. \textit{versions from 1.0.14 to 1.1.4 are deprecated}, \textit{(Another project) seems unmaintained}).
To tackle with the unprecedented size of the dataset, we use a combination of manual and machine-learning labeling to classify real deprecations from candidates.


\begin{figure}
    \centering
    \includegraphics[width=0.85\linewidth]{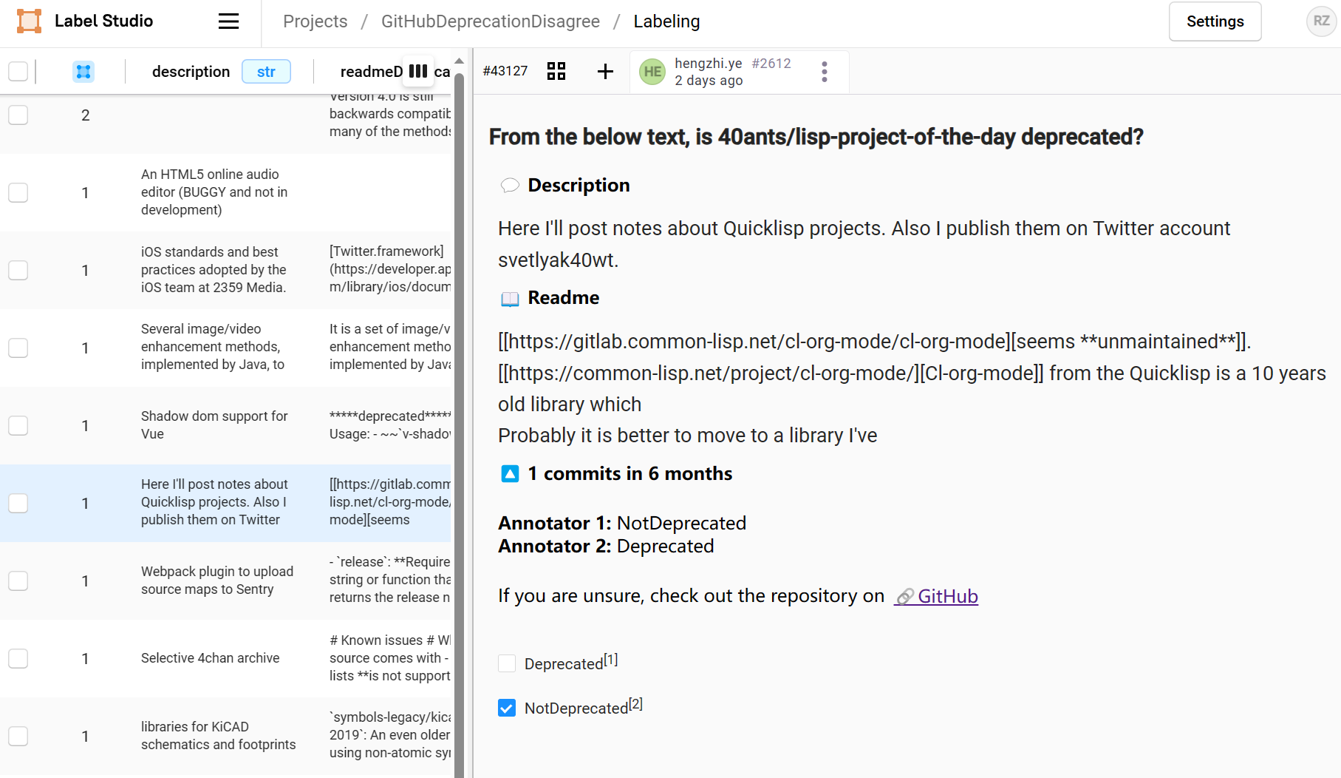}
    \vspace{-.5em}
    \caption{Data Labelling WebUI}
    \label{fig:label_studio}
    \vspace{-1em}
\end{figure}

To ease the labeling experience and make results easily comparable, we first build a delicate Web UI (Figure~\ref{fig:label_studio}) for binary classification tasks with Label Studio~\cite{label_studio}. Two authors independently examine the randomly sampled 1,200 repositories with their commits, issues and pull requests, 
and identify whether the repository has been deprecated or the text indicates a false positive.
The first round of labeling ended with an agreement of 95\% and a Cohen's kappa of 0.814, which suggests a strong agreement. The two authors discuss the inconsistencies and reach a consensus.

Next, we leverage the best-in-class few-shot text classification framework \textit{setfit}~\cite{tunstall2022efficient} to label the remaining 20,358 samples. Setfit is a prompt-free fine-tuning framework for sentence transformers~\cite{reimers2019sentence}, and it is known for its efficiency and accuracy with a small number of labelled samples, which renders it the ideal choice in this scenario.
We fit our binary classifier from the pretrained \textit{paraphrase-mpnet-base-v2}~\cite{hf_mpnet} sentence transformer model on 600 random labelled samples, and use the other 600 as the test set. Our model reaches a stunning accuracy of 0.96, a recall of 0.96 and a precision of 0.90.

After aggregating and deduplicating the two sets, we end up with 51,667 non-fork deprecated GitHub projects.
But those 51,667 projects are not the end of the story. The diversity and representativeness of the dataset is the key to the soundness of the forecast models. Considering that, we incorporate GitHub GraphQL API again to randomly select 51,677 ``known good'' repositories with more than 32 stars and are not identified as deprecated.




\subsection{Dataset Construction}\label{ss:dataprocessing}
Our goal is to forecast the lifespan of GitHub repositories, necessitating the use of time-series data. Hence, the acquisition of historical project statistics, such as the number of stars and commits, is integral to our research. The native GitHub API, however, proves inadequate for this extensive task. Consequently, we turn to the GHArchive dataset~\cite{gharchive}, a more suitable alternative for our purposes.

Unlike other public datasets, which either lack updates or only update annually (e.g. GHTorrent~\cite{gousios2013ghtorent}), GHArchive stands out due to its dynamic nature. It generates fresh data dumps every hour, storing responses from the GitHub TimelineEvent API in a universally compatible JSON format. This dataset provides a comprehensive record of public GitHub activity dating back to 2011, encapsulating various metrics such as pushes, stars, pull requests, issues, and comments.

For the purpose of the analysis, 
we compute the monthly statistics of the projects in our dataset, spanning from 2011 to 2023. To efficiently aggregate these historical metrics from GHArchive dumps, we employ the scalable OLAP database, ClickHouse. Our implementation facilitates rapid retrieval of monthly statistics, even for large-scale projects with over 10,000 issues, achieving sub-second retrieval times.

Eventually we assemble a comprehensive dataset, comprised of 103,307 non-fork projects, spanning from 2011 to 2023.  
Table~\ref{tab:datasets_2} shows the distribution of stars/lifespan of the 51,677 deprecated projects.

\begin{table}[H]
\caption{\textbf{Distribution of star/lifespan of our dataset}}
\centering
\label{tab:datasets_2}
\begin{tabular}{ccccc}
\toprule
&Mean&25\%&50\%&75\% \\
\midrule 
Stars&391.75&48&86&219\\
Lifespan(Days)&1786&978&1673&2243 \\
\bottomrule
\end{tabular}
\end{table}

\section{RQ1: modelling the popularity dynamics of an OSS repository}

In this study we propose that 
a metric modeling the popularity of a repository in its social network on OSS platform is essential to forcast deprecation of an OSS repository. 
Though existing metrics (e.g. commits, stars and release intervals) have been extensively explored and applied to evaluate repositories' activity and the state of maintenance~\cite{mockus2002two, xiaolongterm2023},
those metrics suffer from the following limitations:
1) After years of active development, repositories may start to stabilize and enter a stage with fewer contributions but still consistent maintenance; 
2) Repositories may continue to receive stars after its declaration of deprecation; 
3) Even extraordinarily popular projects may experience a release stall, for example, \textit{react} (a top GitHub project in Table~\ref{tab:top_10_repo}) has not published a new version for 18 months.
The status is calling for a more precise and comprehensive metric. 

Inspired by the interconnected nature and rich historical data of open source communities, we begin with the vast network of repositories and users on GitHub~\cite{blincoe2016understanding},
and propose \textbf{repository centrality}: a family of HITS weight and its normalization on the star bipartite graph of users and repositories.
Below we elaborate on how we define and construct the user-repository network and repository centrality, and conduct preliminary analysis to observe its effectiveness in capturing the complex relationships between repositories and users, as shown in the middle of Figure~\ref{fig:method}.




\subsection{Defining Repository Centrality}\label{modelling-RQ2} 
\begin{figure}
\centering
\includegraphics[scale=0.42,trim=0 0 0 0]{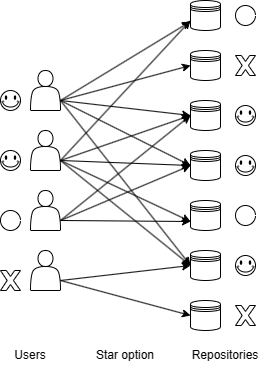}
\vspace{-.5em}
\caption{The Bipartite Graph of Users and Repositories}
\vspace{-.8em}
\label{Repo hits}
\end{figure}

Popularity and attention on an OSS repository is an important indicator of its maintenance status~\cite{crowston2008free}, and it is self-evident that stars are indicators of popularity and attention.
It's also self-evident that just as web pages, stars are not created equal.
Stars from more experienced developers matter more than the ones from newcomers~\cite{kalliamvakou2014promises}.
For instance, stars created by Linus Torvalds are more representative of the preferences of the developer community than that created by novice developers.
We notice that a user can star various repositories, a repository can be starred by distinct users, the ``star'' relationship between users and repositories form a bipartite graph (Figure~\ref{Repo hits}).
There has been extensive effort on node ranking in the field of search engines, and the Hyperlink Induced Topic Search (HITS) algorithm proposed by Kleinberg~\cite{kleinberg1999authoritative,prajapati2012survey} is visible and fits into our context.
HITS assigns scores to nodes based on the link structure of a bipartite graph, nodes with more links to higher importance neighbors are scored higher.

\subsection{Calculation and Normalization}\label{weight-RQ2}

The the HITS algorithm can be formalized as follows: 

\begin{equation}
        Auth(p) = \sum_{q\in p_{to}} Hub(q)
        \nonumber
    \end{equation}
    \begin{equation}
        Hub(p) = \sum_{p\in q_{from}} Auth(p)
        \nonumber
    \end{equation}

The weights of hubs is the aggregation of the weights of its authority neighbors, and vise versa.
In the calculating process, initial weights of hubs and authorities are set to 1, and the algorithm iteratively aggregates the weights, which can be formalized as 
Algorithm~\ref{alg_hits}.

\begin{algorithm}
    \renewcommand{\algorithmicrequire}{\textbf{Input:}}
	\renewcommand{\algorithmicensure}{\textbf{Output:}}
	\caption{Calculation of Repository Centrality} 
    \label{alg_hits}
    \begin{algorithmic}
        \STATE Initialization: $Auth[v] = 1$, $Hub[v] = 1$, $\forall v\in V$
        \REPEAT
        \FOR{each $v$ in $V$} 
        \STATE $Auth[v] = 0$
        \FOR{each $u$ in $incoming\_neighbor(v)$}
        \STATE $Auth[v] = Auth[v]+Hub[u]$
        \ENDFOR
        \ENDFOR
        \FOR{each $v$ in $V$}
        \STATE $Hub[v] = 0$
        \FOR{each $w$ in $outcoming\_neighbor(v)$}
        \STATE $Hub[v] = Hub[v]+Auth[w]$
        \ENDFOR
        \ENDFOR
        \FOR{each $v$ in $V$}
        \STATE $Auth[v] = \frac{Auth[v]}{\sqrt{\sum_{u\in V} Auth[u]^2}}$
        \STATE $Hub[v] = \frac{Hub[v]}{\sqrt{\sum_{u\in V} Hub[u]^2}}$
        \ENDFOR
        \UNTIL convergence
        \ENSURE  $Anth[v]$, $Hub[v]$, $\forall v\in V$
    \end{algorithmic}
\end{algorithm}

The algorithm is relatively simple, 
yet applying it to a gigantic graph of 200 million nodes and 200 million edges is indeed challenging, not mentioning that we need to repeat this process for each month in the past decade.
In face of this challenge, we spent hundreds of developer-hours to implement a efficient and scalable HITS-computation framework. As illustrated in Figure~\ref{fig:method}, our framework digests data from the distributed SQL-compliant database TiDB, and dispatches the workload to an array of Spark nodes.
The above framework allows us to calculate the monthly HITS weight for every GitHub repository over a decade, within a computational budget of 1000 core-hours.


On the other hand, HITS weights, like many metrics in software engineering~\cite{goeminne2011evidence, zhang2021companies}, 
follow a long-tail distribution. The rapid growth of open-source communities has led to a substantial increase in the number of nodes and edges, resulting in significant variance in HITS weights over time. 
Following in the idea of direct comparison of HITS weights across time periods maybe not meaningful,  we apply two normalization methods to stabilize and flatten the HITS weights distribution:

\begin{itemize}
    \item \textbf{Rank normalization:} Following the approach of Mujahid et al.~\cite{mujahid2021toward}, we normalize the HITS weight as its percentile at a given time. The process can be formalized as follows:
    \begin{equation}
        w_\% = \frac{rank(w_i)}{|i|}
        \nonumber
    \end{equation}
\end{itemize}
\begin{itemize}
    \item \textbf{Z-score normalization:} A common linear normalization method that adjusts the values' mean to 0 and standard deviation to 1. Given the long-tail distribution of HITS, we apply a logarithmic transformation to the HITS weight before calculating the z-score:
    \begin{equation}
        w_z = \frac{lnw_i-\mu}{\sigma}
        \nonumber
    \end{equation}
\end{itemize}

We define a repository's centrality to be the combination of its HITS weight in the star bipartite graph, rank-normalized HITS weight, and z-score normalized HITS weight. This composite measure provides a more nuanced understanding of a repository's position within the open-source ecosystem.

\subsection{Preliminary Analysis}
To substantiate our initial hypothesis regarding the correlation between a project's popularity and its lifespan, we perform a preliminary analysis of the collected repositories with the HITS weight calculated 
from the following three aspects. 

\subsubsection{Top-Repositories}\label{ss:top_repo}
Many have found that the number of stars is not a honest representative of a project's quality and impact. An intuitive example is that more often than not, top repositories in GitHub rankings are tutorials, examples and resource collections rather than libraries or frameworks. We find that less expected names of repositories not strongly associated with software development became popular on the GitHub stars leaderboard of September 2023 (Table~\ref{tab:top_10_repo}), for example, 996.ICU, a protest against overwork within the IT industry. In fact, nine of the top ten repositories ranked by the number of stars are not software projects (libraries or applications), the only exception being \textit{react}, a web framework.
From a software developer's perspective, the top repositories ranked by HITS sound more familiar and influential: top ten repositories include seven well-known software projects like \textit{tensorflow}, \textit{vue} and \textit{vscode}.
This intuitive observation confirms on the HITS weight's capability of comprehending repository impact, which in turn influences repository lifespan~\cite{ait2022empirical}. 





\begin{table}[H]
\vspace{-1.2em}
\caption{\textbf{Top 10 Repositories ranked by HITS and Stars}}
\centering
\label{tab:top_10_repo}
\begin{tabular}{ccc}
\toprule
Rank & Repository by HITS & Repository by Stars \\
\midrule 
1 & sindresorhus/awesome & freeCodeCamp \\
2 & facebook/react*  & free-programming-books \\
3 & vuejs/vue* & sindresorhus/awesome \\
4 & tensorflow* & 996.ICU \\
5 & airbnb/javascript* & coding-interview-university \\
6 & You-Dont-Know-JS & public-apis \\
7 & react-native* & developer-roadmap \\
8 & oh-my-zsh* & system-design-primer\\
9 & developer-roadmap & build-your-own-x\\
10 & Microsoft/vscode* & facebook/react* \\
\bottomrule
\end{tabular}
\\
\vspace{.5em}
\RaggedRight \quad \quad { \footnotesize* Software projects (applications or libraries).}
\vspace{-1em}
\end{table}


\subsubsection{Representative Projects}\label{ss:case_study}

How does HITS work in practice? How is its capability of forecasting deprecation compared with other metrics? 
To answer the questions, firstly we choose the once-popular code editor Brackets, which was deprecated in favor of the more feature-complete VSCode in 2021, as a showcase. 
Secondly we 
explore the prediction power of HITS's delta $\Delta$HITS ($\Delta HITS = HITS_{t} - HITS_{t-1month}$)
within three randomly sampled projects: Project~\cite{0age}, ~\cite{0mniscient} and~\cite{00-Evan}.



\begin{figure}
    \centering
    \includegraphics[scale=0.42,trim=0 0 0 0]{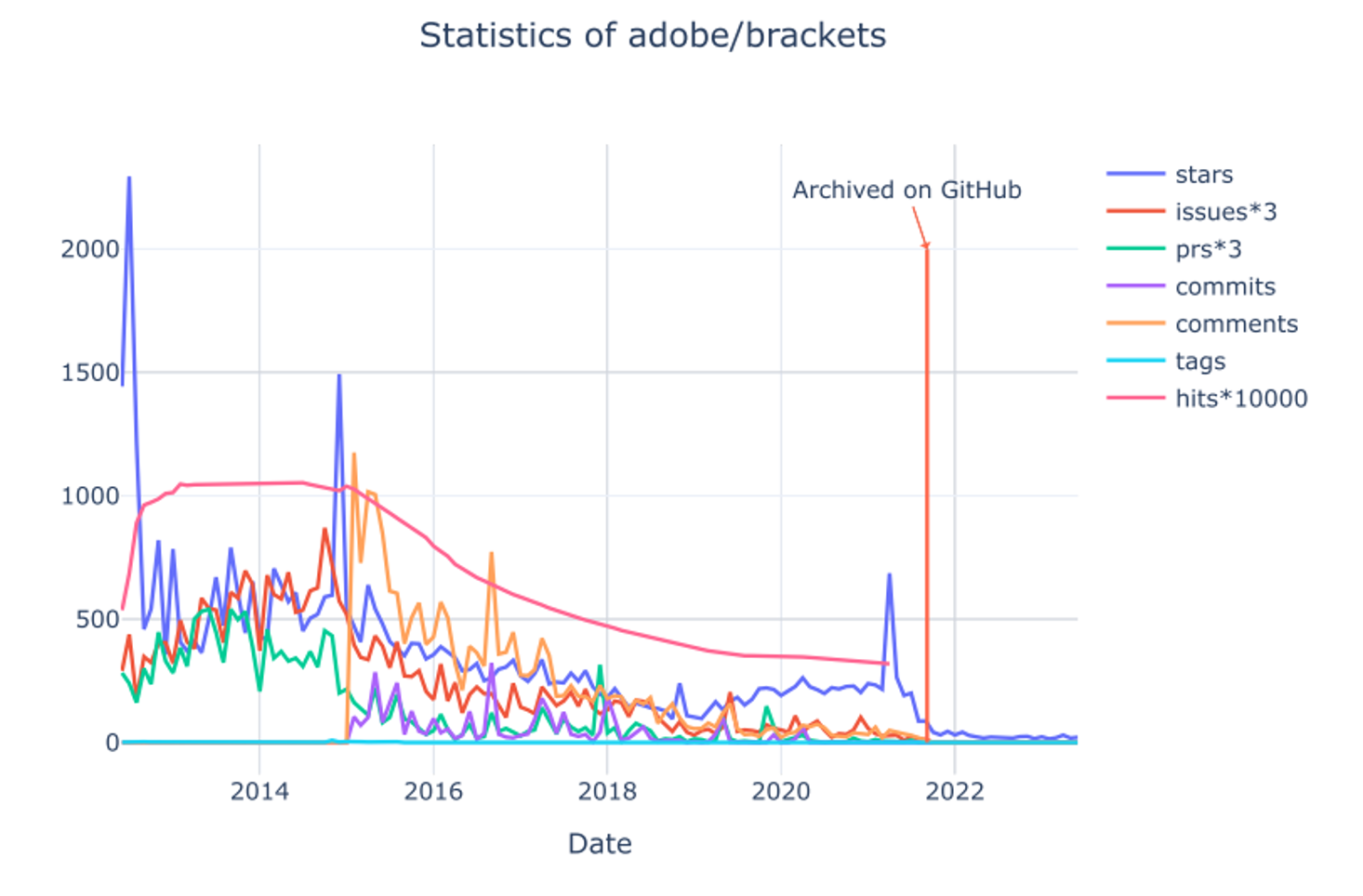}
    \vspace{-1em}
    \caption{Repository features over time of Adobe/Bracket}
    \label{feature_trends}
    \vspace{-1em}
\end{figure}

Figure~\ref{feature_trends} displays the activity statistics of the Brackets project since 2015. The figure shows the number of new events created each month for stars, issues, PRs, commits, comments, and tags. It is evident that the development of Brackets has been gradually stagnating since 2015. Yet, the project continued to receive a high and stable number of stars each month, with a notable surge in 2021. The HITS weight however, as a reliable indicator of the project's impact, has been on a steady decline since 2015.
This case demonstrates the HITS weight to be a more promising representation of projects' deprecation trends, and less prone to noise compared with other indicators.


\begin{figure}
    \centering
    \subfigure[HomeWork\vspace{-1em}]{
        \centering
        \includegraphics[width=0.45\linewidth, trim=0 0 0 20, clip]{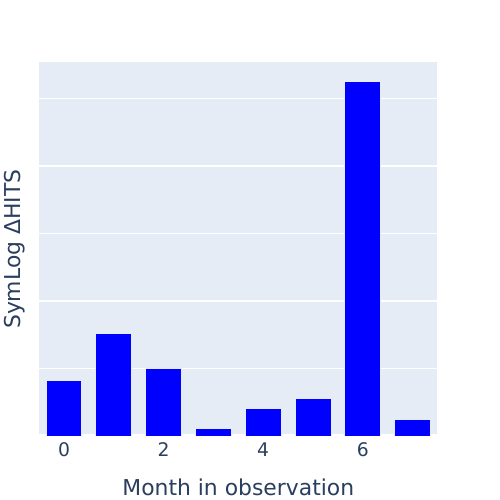}
        \label{0age}
    }
    \subfigure[Discord-Themes\vspace{-1em}]{
        \centering
        \includegraphics[width=0.45\linewidth, trim=0 0 0 20, clip]{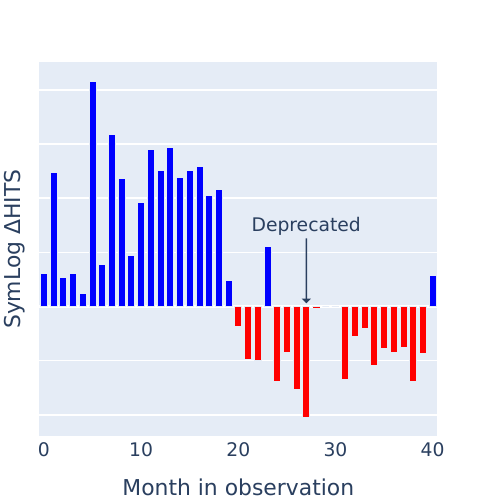}
        \label{0mniscient}
    }
    \subfigure[shattered-pixel-dungeon-gdx]{
        \vspace{-1em}
        \centering
        \includegraphics[width=0.45\linewidth, trim=0 0 0 20, clip]{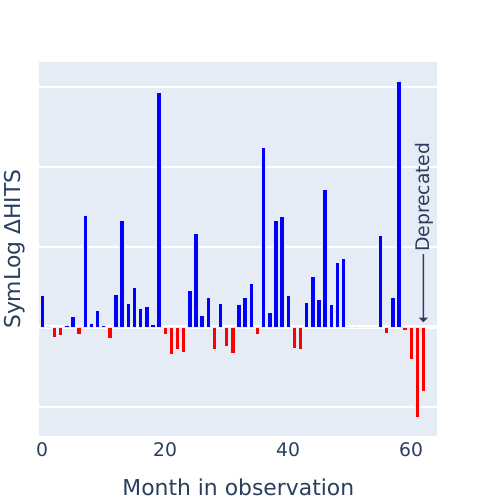}
        \label{00-Evan}
    }
    \vspace{-.5em}
    \caption{$\Delta$HITS over time of three open-source projects}
    \label{delta_hits}
    \vspace{-1.5em}
\end{figure}

Figure~\ref{delta_hits} illustrates the relationship between $\Delta$HITS and time for the three randomly selected projects. It is clear that for Project~\cite{0age}, represented in Figure~\ref{0age}, there was no negative $\Delta$HITS during the observation period. Indeed, this project has not been deprecated and is still under active maintenance. However, for Project~\cite{0mniscient} and~\cite{00-Evan}, represented in Figures~\ref{0mniscient} and~\ref{00-Evan} respectively, deprecation occurred during the observation period. In each case, a negative peak in $\Delta$HITS in the month preceding deprecation serves as a harbinger of this event.

Therefore, it is evident that the HITS weight, as a predictor of repository deprecation, exhibits a high degree of sensitivity and have the potential to accurately detect a trend towards deprecation.

\subsubsection{Correlation Between HITS metric and Other Metrics}

Is HITS truly unique? Does it capture trends that metrics developers already rely on, such as the number of stars and commits, fail to reveal? 
To get some answers,
we follow the methodology of previous work~\cite{mujahid2021toward} and select Spearman's rank correlation test~\cite{kendall1938new} to measure the correlation between HITS and other metrics. Spearman's rank correlation coefficient ($\rho$) is calculated from the relative order of the value, and ranges from -1 and 1, with -1 indicating perfect negative correlation and 1 vise versa.  Spearman's $\rho$ is ideal in this context for its ability to mitigate disturbances caused by data distribution.

\begin{figure}
    \centering
    \subfigure[Correlation Matrix]{
        \centering
        \includegraphics[width=0.5\linewidth,trim=5 32 -2 30,clip]{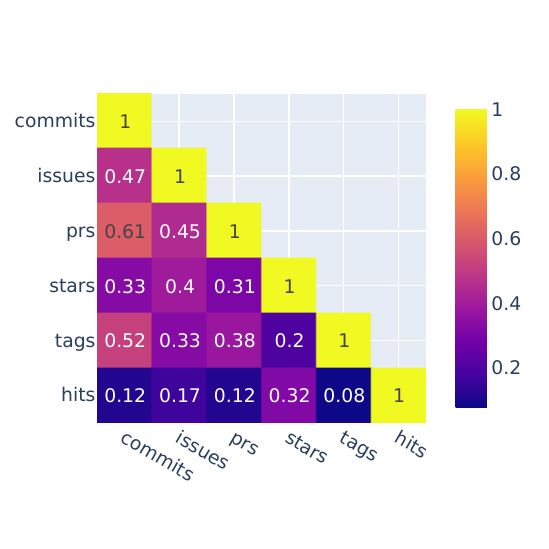}
        \label{fig:corr_matrix}
    }
    \subfigure[Correlation by Repository]{
        \centering
        \includegraphics[width=0.4\linewidth,trim=25 20 10 30,clip]{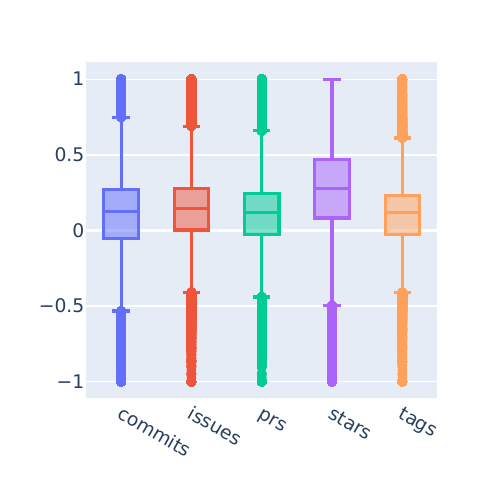}
        \label{fig:corr_violin}
    }
    \vspace{-.5em}
    \caption{Spearman's Correlation Between Features}
    \label{fig:corr_res}
    \vspace{-1.8em}
\end{figure}

Figure~\ref{fig:corr_matrix} presents the correlation matrix between metrics. We observe that Spearman's $\rho$ between HITS and other metrics across all samples is less than 0.4, suggesting negligible to weak correlations, as interpreted by Prion and Haerling~\cite{prion2014making}. The sole strong correlation identified is between the number of pull requests and the number of commits.
Further analysis of the distribution of correlations among repositories is provided in Figure~\ref{fig:corr_violin}, from which we can draw the following conclusions: 1) In most repositories, HITS is positively correlated with the number of commits, issues, PRs, stars, and tags, which is reasonable; 2) The 75th percentiles (the upper edge of the boxes in the figure) are all below 0.4, indicating that HITS exhibits weak correlations with the metrics in the majority of repositories. The only exception is the moderate correlation between HITS and the number of stars. 
The results 
indicate that HITS may capture unique characteristics of the repositories that can not be simply modelled by existing common metrics.

\begin{result-rq}{Summary for RQ1:} 
We craft a comprehensive repository centrality metric that captures the dynamics of popularity of open-source projects from the bipartite star network between users and repositories. 
Preliminary analysis confirms its effectiveness and uniqueness in representing project popularity, offering a promising entry point for predicting the lifespans of OSS repositories.
\end{result-rq}

\section{RQ2: Effectiveness of repository centrality}




To estimate the prediction power of repository centrality, we turn to 
an approach that is often employed to estimate the survival hazard: survival analysis.
The life cycle of an OSS project repository commences upon its creation, with various activity metrics being recorded until its deprecation, which could be compared to that of a human. The activity data of the repository (e.g., stars and commits) is similar to the vital signs of a human (e.g., blood pressure and glucose), and the survival time of the repository is similar to a human's life expectancy. Survival analysis, a prominent method in medical research 
can be aptly applied here. As a form of regression that models hazard rate over time of an event, adept at handling censored data, it matches our data and prediction target well.
Note that studies in software engineering have widely adopted survival analysis as well~\cite{survival2010,zhou2016inflow}.

In particular, we fit survival analysis models with the repository centrality metric and other historical metrics being predictors, to predict the survival time (AFT) or hazard rate (DRSA) of a repository from the current observation point, as shown on the right of Figure~\ref{fig:method}.

\subsection{Selection of Controlling Features}\label{predictivemodel-RQ3}
 
Our goal is to train and fit suitable models based on 
 HITS weight and various controlling features, to predict the expected lifespan of a repository from the current observation point. 

To offer comprehensive baselines to repository centrality being predictor, 
we extract repository features (controlling features) from three perspectives:

\begin{itemize}
    \item \textbf{Development:} The number of new commits serves as a robust indicator of a project's development and maintenance activity. A near-deprecation project tends to exhibit a steep decline in the frequency of bug fixes and features, which means a stall in commits and releases.
\end{itemize}
\begin{itemize}
    \item \textbf{Collaboration:} Sustained work within a core group of contributors is a significant contributor to the success and longevity of open-source projects~\cite{joblin2022successful}. 
    On code hosting platforms like GitHub, developers collaborate by opening issues, pull requests and discuss under comments. They are good proxies of a project's collaboration status.
\end{itemize}
\begin{itemize}
    \item \textbf{Community Attention:} The number of stars is a widely used popularity metric in software engineering studies. 
    As a complement to the number of stars and the core contribution of this work, we introduce repository centrality (HITS weight, rank-normalized HITS, and z-score normalized HITS) as encodings of a project's community attention.
\end{itemize}

The features we ultimately select as controlling predictors are listed in Table~\ref{tab:selected_features}.

\begin{table}[H]
\vspace{-1em}
\caption{\textbf{Selected Features}}
\centering
\label{tab:selected_features}
\begin{tabular}{cc}
\toprule
Feature&Explanation \\
\midrule 
Commits&Created Commits in a Month\\
Comments&Created Issue / PR Comments in a Month \\
Issues&Created Issues in a Month \\
PRs&Created PRs in a Month \\
Stars&New Stars in a Month \\
Tags&Created Tags in a Month \\
Weight&HITS Weight of the Project Repository \\
Weight$_\%$&Percentile Normalized HITS \\
Weight$_z$&Z-Score Normalized HITS \\
\bottomrule
\end{tabular}
\vspace{-.5em}
\end{table}


\subsection{Accelerated Failure Time (AFT)}\label{suvival-RQ3}
The Accelerated Failure Time (AFT) model proposed by Fox et al.~\cite{cox1972regression}, is one of the most commonly used models in survival analysis. It assumes a linear relationship between the logarithm of survival time and the predictor variables, expressed as: $lnY = \left< w, x\right> + \sigma \cdot Z$, where $x$ is a vector representing the features, $w$ is the coefficient vector, $Y$ is the output label (survival time), $Z$ is a known probability distribution of noise, and $\sigma$ is the scaling factor of $Z$. In this study, $x$ represents the array of features of GitHub projects, and the label to predict, $Y$, is the survival time of the projects.

To evaluate the practical utility of the HITS weight, we train an AFT model using the state-of-the-art XGBoost~\cite{barnwal22survival} gradient boosting framework based on the selected features, including HITS weights.
XGBoost has demonstrated top-tier performance in numerous prediction tasks and exhibits impressive scalability.
It is also easily interpretable with the reported F-Score and automatic feature selection.

We set aside 20\% of the repositories as the test set for the model, with nloglik chosen as the loss function. To accelerate the training process, we limited the iteration rounds to 50 and set the number of early-stopping rounds to 10. Other parameters were set to XGBoost defaults.




This model converged after 50 iterations, and the C-Index on the test set was 0.810. This score indicates that the model has a strong discriminatory power~\cite{li2017nomograms}. 

Thanks to the interpretable nature of gradient boosting models, we can derive quantitative insights into how well the HITS weights contribute to the predictions. First, we choose the F-Score calculated by the XGBoost framework as the direct measurement of the features' contributions;
Besides, following the choice of numerous AI researchers~\cite{borisov2022deep}, we pick SHAP (SHapley Additive exPlanations)~\cite{lundberg2017unified} as the generic metric of feature importance. By borrowing concepts of ``cooperator'' and ``payoff'' from the game theory, SHAP calculates the average contribution of each feature to the model output delta by considering all possible feature combinations.

\begin{figure}
    \centering
    \subfigure[XGBoost's F-Score]{
        \centering
        \includegraphics[width=0.9\linewidth]{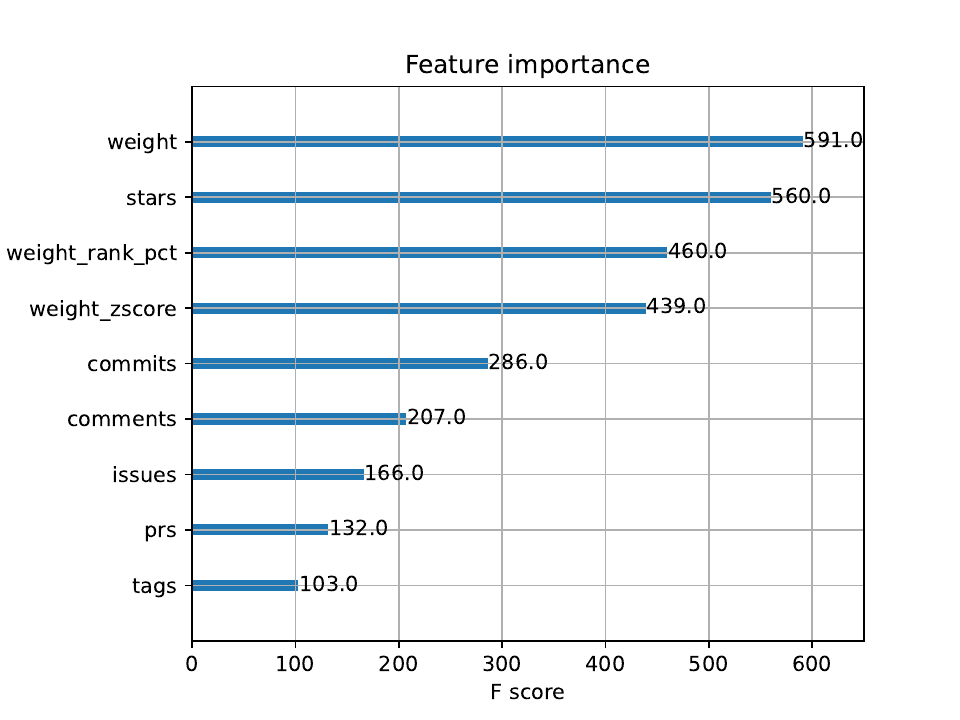}
        \label{fig:xgb_feature_imp}
    }
    \subfigure[Beesworm Diagram of Shapley Values]{
        \centering
        \includegraphics[width=0.92\linewidth]{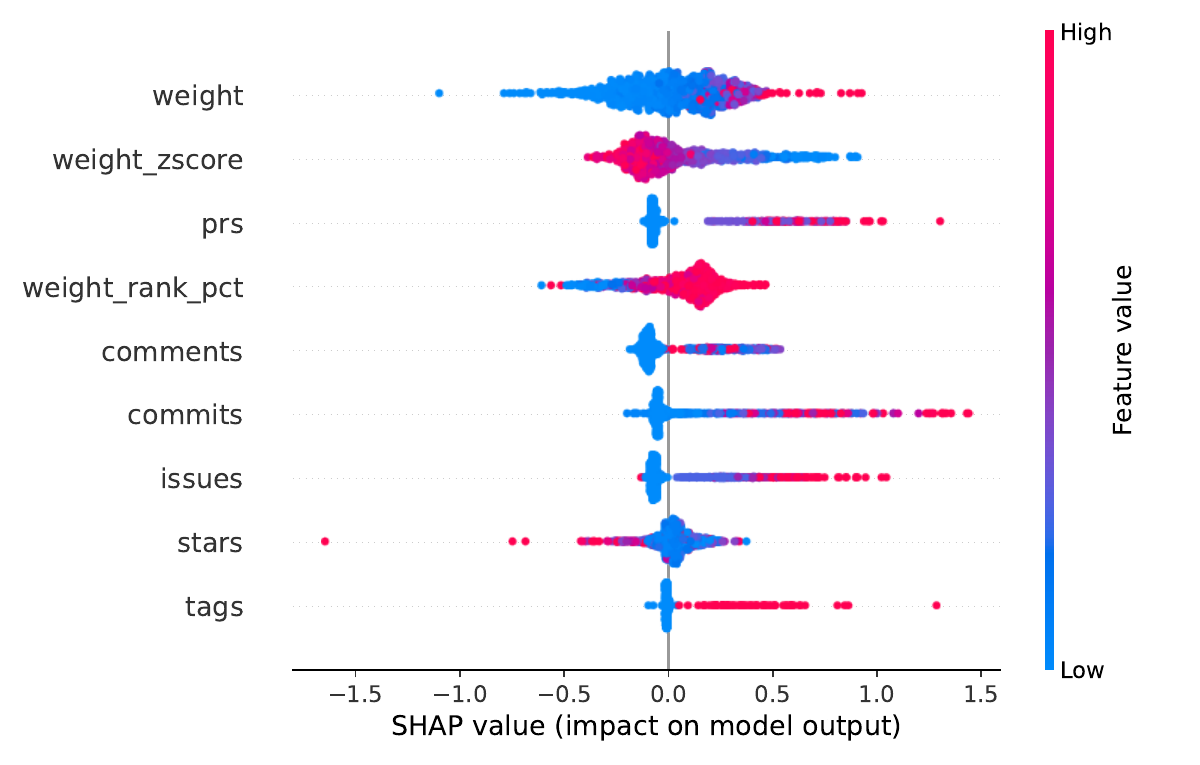}
        \label{fig:xgb_shap}
    }
    \caption{Feature Importance of the XGBoost AFT Model}
    \vspace{-.5em}
    \label{fig:xgb_perf}
    \vspace{-.5em}
\end{figure}

Figure~\ref{fig:xgb_feature_imp} presents the F-scores of each feature in the XGBoost model, showing that the HITS weight has the highest importance, followed by the number of stars. Percentile-normalized HITS and Z-score normalized HITS ranked third and fourth, respectively, suggesting that decreasing popularity significantly influences developers' decisions to deprecate repositories, while the frequency of maintenance and collaboration has a less impact. 
Figure~\ref{fig:xgb_shap} shows that the HITS weight and Z-Score normalized HITS have highest SHAP values, suggesting that they have strongest effect on the predictions. The red dots on the right side and the blue cluster on the left side indicate a positive correlation between HITS weight and prediction values. In other words, projects with higher HITS weights have a greater chance to survive.

\begin{table}
    \centering
    \caption{\textbf{Performance of Ablation Models (AFT)}}
    \vspace{-.5em}
    \label{tab:xgb_ablation_perf}
    \begin{tabular}{lcc}
        \toprule
        \textbf{Model} & \textbf{C-Index} & \textbf{Predicted Lifespan (Days, Mean)} \\
        \midrule
        Baseline        & 0.748 & 7877 \\
        Baseline-stars  & \underline{0.718} & 7925 \\
        Full-weight     & 0.755 & \underline{7663} \\
        Full-weight$_\%$ & 0.808 & \textbf{7926} \\
        Full-weight$_z$  & 0.808 & 7726 \\
        Full-comments   & 0.805 & 7772 \\
        Full-commits    & 0.801 & 7685 \\
        Full-issues     & 0.805 & 7569 \\
        Full-prs        & 0.804 & 7374 \\
        Full-stars      & 0.809 & 7877 \\
        Full-tags       & 0.809 & 7914 \\
        Full            & \textbf{0.810} & 7925 \\
        \bottomrule
    \end{tabular}
    \vspace{-.5em}
\end{table}

Table~\ref{tab:xgb_ablation_perf} illustrates the results from the ablation study. The inclusion of HITS has a noticeable effect on the performance of the predictor. Compared to the baseline model, the full model shows a significant improvement in C-Index, from 0.748 to 0.810. 
Removing the HITS weight feature from the full model results in a drop in the C-Index to 0.755, while no significant drop was observed after removing other features. This suggests that HITS weights plays a unique role in lifespan prediction for GitHub projects. Given that HITS weights does not nullify the overall evolution in GitHub projects, it represents an important avenue for future researchers to explore the evolution of project deprecation.


\subsection{Deep Recurrent Survival Analysis (DRSA)}\label{neural-RQ3}

We also employ the Deep Recurrent Survival Analysis (DRSA) proposed by Ren et al~\cite{ren2019deep} to predict repository deprecation with the memory of historical shifts of the features. 
As a survival analysis model based on recurrent neural networks (RNNs), DRSA is good at capturing trends of metrics in the timeline
and is perfect for the job.

We randomly select 80\% of the repositories in our dataset as the training set, with the remaining 20\% serving as the test set. For each repository, we extract each month’s features from the historical data over a continuous 10-month period and then feed the sequential data into the DRSA model for training. We set batch size to 64, learning rate to 0.015 and iteration rounds up to 1000 to conduct the training process.

\begin{figure}
    \centering
    \subfigure[SwiftEducation/Stopwatch]{
        \centering
        \includegraphics[width=0.9\linewidth]{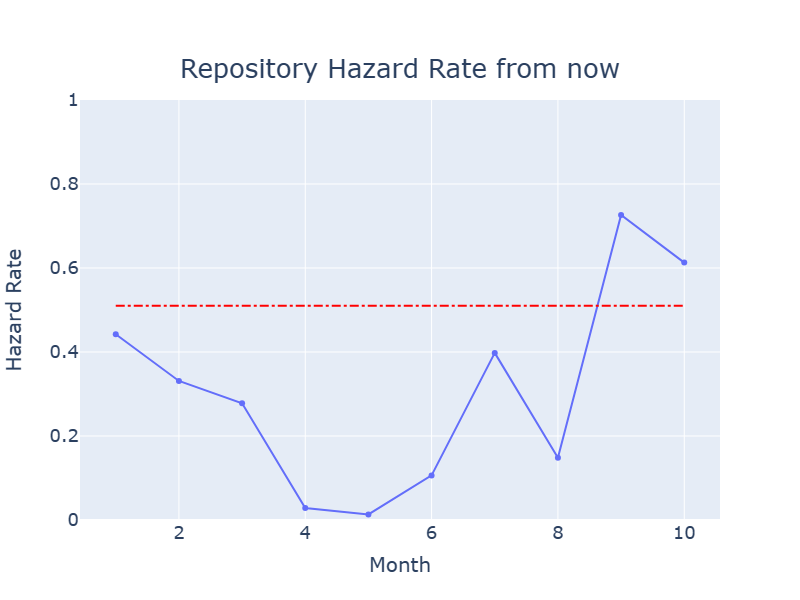}
        \vspace{-.5em}
        \label{fig:stopwatch}
    }
    \subfigure[google/mr4c]{
        \centering
        \includegraphics[width=0.9\linewidth]{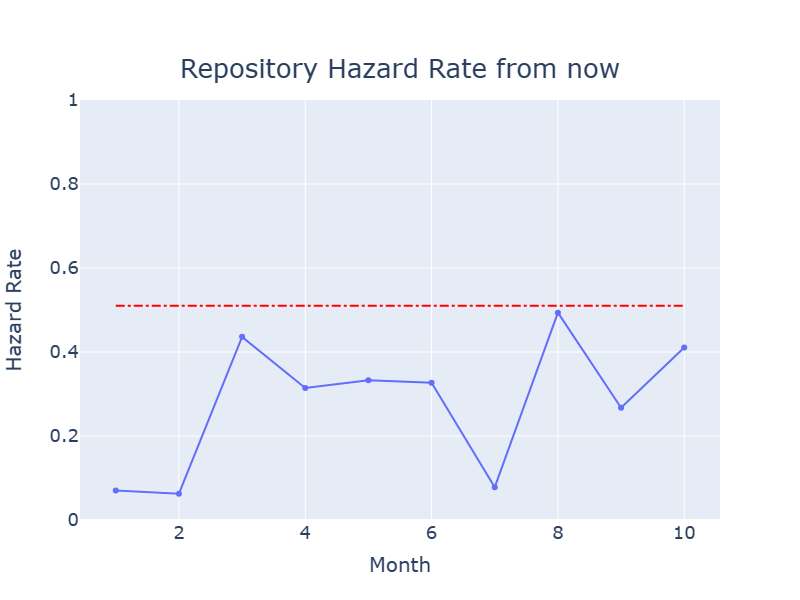}
        \vspace{-.5em}
        \label{fig:mr4c}
    }
    \caption{Two samples of the DRSA prediction results}
    \vspace{-0.5em}
    \label{fig:drsa_sample}
    \vspace{-1em}
\end{figure}

After training, we evaluate the DRSA model on the test set and get satisfactory results.
We randomly select two repositories Stopwatch~\cite{Stopwatch} and mr4c~\cite{mr4c} from the test set to illustrate the prediction ability of DRSA models trained on monthly HITS weight and other metrics for repository prediction. As shown in Figure~\ref{fig:drsa_sample}, the horizontal coordinate indicates the number of months since our last observation, while the vertical coordinate indicates the hazard rate of repository depreciation in each month. We assume that a hazard rate greater than 50\% indicates a high probability of repository depreciation in the corresponding month, which is a wake-up call that deprecation may occur. Meanwhile, if the line graph shows a peak of more than 50\% in the observation interval covered by the horizontal coordinate, we intuitively consider the month corresponding to the peak as the month in which the DRSA model predicts the occurrence of deprecation.

In Figure~\ref{fig:stopwatch}, we use the feature data of Stopwatch from January 2017 to October 2017 as a test input, and the hazard rate curve shows the highest peak probability of deprecation (73\%) in the ninth month ahead, and we assume that the project will be discarded in the next nine months. According to the actual situation, the project was archived on 5 July 2018. The difference between the result we predicted based on the model and the real deprecated date is within one month, and the predicted date is earlier than the actual deprecated date, so we can consider that we have a certain degree of accuracy on the prediction. While in Figure~\ref{fig:mr4c}, we use the data of mr4c from October 2015 to July 2016 as the test input, and the hazard curve does not exceed 50\% in any of the next 10 months, which can be considered that the project will not be deprecated in the short term. In fact, mr4c was archived in December 2021, well after 2016, and therefore matches the prediction.

Similar to the ablation study results of AFT model, the inclusion of HITS strongly contributes to the performance of the DRSA model: the full model shows a significant improvement in C-Index of 9.3\% than base model, and removing the HITS weight related features from the full model results in a drop in the C-Index of 6.9\% while less significant drop (less than 4\%) was observed if other features removed. These results further support that HITS weights are of good use in lifespan prediction for GitHub projects.

\begin{result-rq}{Summary for RQ2:} 


We utilize the latest developments in gradient boosting and deep learning to fit survival analysis models on GitHub repositories, with both XGBoost AFT and DRSA models excelling in the C-Index metric. Ablation studies confirm that HITS weights are the most critical factors influencing the results. The performance metrics and ablation studies for the models highlight the remarkable predictive capabilities of HITS weights.
\end{result-rq}

\vspace{.5em}

\section{Discussion}\label{discussion}
In this section, we discuss how our approach may help with related research and practice, and what factors may potentially threaten the validity of this study.

\subsection{Implication to Research and Practice}
\textbf{Novel Use of Centrality Indicator.} In contrast to traditional metrics, our study introduces a centrality measure based on the HITS algorithm, providing a more nuanced depiction of repository popularity. This innovative approach encourages a higher-dimensional perspective on the issue of repository deprecation trends. Our work aligns with the increasing recognition in 
the importance of network-related indicators in capturing the evolving nature of OSS development ecosystems. Moreover, it reveals a novel angle for analyzing complex system through network analysis, echoing recent advances in network analysis applications to software engineering research.

\textbf{Scalable Approach.} 
Our approach leverages graph computation acceleration tools, enabling efficient calculation of HITS weights and potential scalability to larger datasets, given sufficient computational resources. This aligns with the broad trend towards the use of scalable methodologies in software engineering research to tackle large-scale problems. While our study focuses on GitHub, the proposed methodology could be applicable to other open-source platforms, offering a versatile framework for predicting repository deprecation risks. Furthermore, our predictive model, based on HITS weight and survival analysis model, performs well in forecasting deprecation risks, providing a foundation for future research in this area. 

\subsection{Threats to Validity}
\textbf{Internal Validity.} The first threat pertains to the use of the HITS algorithm, which favors older, more established projects due to its recursive nature~\cite{kleinberg1999authoritative}. This could potentially introduce a bias towards these projects. 

The second threat concerns the survival analysis fitting process. While we have considered a set of features, there might be other influential factors not included in our model or feature overlap existing. Furthermore, the performance of the survival analysis model could be sensitive to the choice of hyperparameters during the training process, which could affect the predictive power of the model~\cite{harrell2001regression}.

In addition, our bipartite graph model only considers the star-relationship between users and repositories. However, users and repositories might be connected through other relationships, such as forks or pull requests, which could yield a more complex network structure~\cite{gousios2014exploratory,badashian2014involvement}. And we don't include star cancellations in our dataset, which are not given by the GitHub Timeline API, so there may also be discrepancies between the network we constructed and the real one.

\textbf{External Validity.} Since our research is primarily based on data from GitHub, which has its unique user base and project characteristics, there are potential threats to validity. Though similar, these characteristics may not be representative of other open-source platforms, potentially limiting the applicability of our findings.

Besides, our study dose not take the influence that project type may have on a project's popularity and deprecation risk into consideration. However, it might not hold true in all cases. At a similar degree of centrality, different types of projects (e.g., commercial, volunteer, interest-driven) may have different deprecation risk, which potentially affects the validity of our findings~\cite{capiluppi2007cathedral}.

\section{Conclusion}\label{conclusion}
OSS repositories face the risk of deprecation for various reasons~\cite{coelho2017modern}, leading to instability in the software ecosystem. Therefore, predicting repository deprecation is crucial for ensuring the sustainability of OSS projects.

In this paper, we address this pressing need in the software engineering domain, a gap that existing metrics have failed to bridge adequately. Traditional metrics, while informative, do not fully capture the network centrality of OSS project popularity and its implication for repository deprecation.

To bridge this gap, we propose and validate a centrality metric based on the HITS algorithm, which captures the repository popularity according to the connections between users and repositories in a bipartite network. Our method provides a more comprehensive understanding of OSS project popularity and its correlation with repository deprecation. Our preliminary analysis and subsequent survival model analysis demonstrate the effectiveness of the HITS weight in predicting repository deprecation, with our models showing good performance.

Looking forward, there is potential for further refinement and validation of our metric 
across various open-source platforms. Moreover, the integration of our metric into predictive models could be explored to enhance the accuracy of deprecation forecasts. Such advancements would not only provide developers and maintainers with a more nuanced understanding of project popularity and repository deprecation but also equip them with a reliable tool for strategic decision-making, thereby fostering a more sustainable open-source software ecosystem.


\bibliographystyle{IEEEtran}
\bibliography{references}

\end{document}